\documentclass[aps,prb,twocolumn,groupedaddress,showpacs]{revtex4}

\usepackage{graphicx} % uncomment this line to include the graphicx package
\usepackage{amssymb} % for the square root function

\usepackage{amsmath} %follow EK
\usepackage{dcolumn} %follow EK
\usepackage{bm}      %follow EK

\begin{document}

%\preprint{}

\title{Thermal History of Solid $^4$He Under Oscillation}


\author{A. C. Clark}
    \email{cctony1@gmail.com}
\author{J. D. Maynard}
\author{M. H. W. Chan}
%\thanks{}
\affiliation{Department of Physics, The Pennsylvania State
University, University Park, Pennsylvania 16802}


\date{\today}

\begin{abstract}

We have studied the thermal history of the resonant frequency of a
torsional oscillator containing solid $^4$He. We find that the
magnitude of the frequency shift that occurs below $\sim$100 mK is
multivalued in the low temperature limit, with the exact value
depending on how the state is prepared. This result can be
qualitatively explained in terms of the motion and pinning of
quantized vortices within the sample. Several aspects of the data
are also consistent with the response of dislocation lines to
oscillating stress fields imposed on the solid.

\end{abstract}

%\keywords{}

\maketitle

\section{\label{sec:intro}Introduction}

Over the last several years the torsional oscillator (TO) technique
has become a popular method
\cite{nature,science,jltp,prl,rr1,rr2,kojima,kubota,shirahama,tj,he3,penzev}
to study solid $^4$He. The measurements, which involve monitoring
the resonant frequency \textit{f} of a high quality factor torsion
pendulum filled with $^4$He, have revealed a ``transition'' below an
onset temperature \textit{T$_O$} of $\sim$200 mK. The results of
several control experiments \cite{science} indicate that the sudden
increase in the frequency is due to a superfluid-like decoupling of
a fraction of the $^4$He mass. The apparent nonclassical rotational
inertia fraction (NCRIF), proportional to the frequency shift
$\delta$\textit{f}, is independent \cite{science} of the maximum
oscillation speed \textit{v}$_0$ of the TO provided that it is
smaller than a critical value. Above this critical velocity
\textit{v$_C$} the magnitude of NCRIF is attenuated. The value of
\textit{v$_C$} corresponds to several quanta of circulation
$\kappa$, suggesting that the important excitations in the system
are vortices. Anderson has proposed \cite{anderson} a vortex liquid
model that qualitatively captures a number of experimental results.

In multiple solid $^4$He samples slowly grown within an annulus of
radius, \textit{r} = 5 mm, and width \cite{error}, \textit{t} = 0.95
mm, a value of \textit{v$_C$} = 10 $\mu$m$\,$s$^{-1}$ was found
\cite{prl} (corresponding to $\kappa$ = 2$\pi$\textit{rv}$_0$
$\approx$ 3\textit{h}/\textit{m}, where \textit{h} is Planck's
constant and \textit{m} is the bare $^4$He mass). NCRIF decreased
almost linearly with ln[\textit{v}$_0$] as the speed was increased
further. This attenuation, according to Ref.~\onlinecite{anderson},
is attributable to the nonlinear susceptibility of a vortex liquid
phase, which consists of an entanglement of many thermally activated
vortices \cite{anderson}. The ability of the vortices to move
counter to the time-dependent superflow (relative to the cell's
oscillation) results in the screening of supercurrents. As the
temperature is lowered the motion and number of vortices is reduced
so that NCRIF becomes finite. The accompanying dissipation
\cite{science} that peaks in the middle of the transition is due to
a matching of \textit{f} and the optimal rate at which vortices
respond to changes in the velocity field. Residual $^3$He atoms may
\cite{SaslowWu} cling to vortices, thereby slowing vortex motion and
ultimately enhancing the onset temperature \cite{nature,tj,he3}.

One prediction of Anderson's model \cite{anderson} is that
\textit{T$_O$}, and perhaps NCRIF, for a particular sample will
decrease as the measurement frequency is lowered. Aoki \textit{et
al.} performed \cite{kojima} measurements in a double oscillator
that operated in either an antisymmetric (high \textit{f}) or a
symmetric (low \textit{f}) mode and found that for the same $^4$He
sample, \textit{T$_O$} $\approx$ 240 mK at 1173 Hz and
\textit{T$_O$} $\approx$ 160 mK at 496 Hz. However, NCRIF became
independent of frequency below $\sim$35 mK. In addition,
irreversible changes in NCRIF below $\sim$45 mK occurred upon
variation of the oscillation speed between 10 $\mu$m$\,$s$^{-1}$ and
800 $\mu$m$\,$s$^{-1}$.

%Vortices have long been known \cite{hall} to exist as elementary
%excitations in superfluid liquid $^4$He. A straightforward method
%\cite{hess} to produce quantized vortex lines is by rotating a
%container of liquid $^4$He with a circulation of $\kappa$ $\ge$
%\textit{h}/\textit{m} while cooling through the superfluid
%transition. To remove the vortex from a simply connected system the
%dc rotation speed must be lowered below a threshold value to allow
%its escape (along with a 2$\pi$ phase slip). However, some remnant
%vorticity is often present \cite{remnant} since the ends of vortex
%lines can attach to the container walls. In a variety of other
%experiments it has also been found that vortex lines and rings are
%pinned by nonsuperfluid regions, such as a wire \cite{vinen} (which
%serves as a useful measurement tool of circulation) penetrating a
%bath of liquid $^4$He, porous media \cite{koj}, and impurity atoms
%or ions \cite{ions}.

Nonsuperfluid explanations of the TO experiments have also been
proposed, in which some sort of freezing process \cite{balatsky1} or
glass transition \cite{balatsky2} takes place. Although the former
was only discussed qualitatively, the pinning of dislocation lines
within solid $^4$He is one possible realization of such a
``freezing'' transition. The mechanism for dislocation pinning is
likely related to interactions with isotopic impurities. It is known
that dislocations \cite{wanner} and $^3$He atoms
\cite{lin,impuriton} are mobile in solid $^4$He below $\sim$1 K, but
the strong coupling between them in the low temperature limit can
immobilize them both \cite{iwasa,pbd}. In a TO study \cite{he3} of
many solid $^4$He samples we found that the dependence of
\textit{T$_O$} on the $^3$He concentration \textit{x}$_3$ is
quantitatively consistent with this impurity-pinning model.

Recently Day and Beamish observed \cite{day} a dramatic increase
(between 5\% and 20\%) in the shear modulus \textit{c}$_{44}$ of
solid $^4$He (\textit{x}$_3$ $\approx$ 0.3 ppm) below $\sim$250 mK,
with the temperature dependence of \textit{c}$_{44}$ strongly
resembling that of \textit{f} in TO experiments. The authors also
reported hysteresis in \textit{c}$_{44}$ upon adjusting the stress
amplitude of the measurement at \textit{T} $\approx$ 20 mK, similar
to what is seen in Ref.~\onlinecite{kojima}. The increase in the
shear modulus is very likely due to the stiffening of the
dislocation network via the immobilization of individual lines. It
was suggested by the authors of Ref.~\onlinecite{day} that an
increase in \textit{c}$_{44}$ will result in a stronger coupling
between the $^4$He sample and its container and therefore mimic an
enhanced mass loading of a TO, causing \textit{f} to decrease.
However, any enhancement to the overall rigidity of the system
actually leads to a higher resonant frequency. We have carried out a
finite element method (FEM) calculation \cite{FEM} employing
realistic parameters of the TO and solid helium in order to confirm
this result. The positive correlation between \textit{c}$_{44}$ and
\textit{f}, their resemblances in the temperature dependence, and
the hysteresis common to both, all suggest that either the frequency
shift is a direct consequence of the enhanced modulus and can be
explained without invoking supersolidity or that the microscopic
mechanism responsible for NCRI also affects the elastic properties
of the solid (e.g., a recent suggestion is given in
Ref.~\onlinecite{arxiv}). The results presented below appear to
favor the latter.

\begin{figure}[t]
\includegraphics[width=1.0\columnwidth]{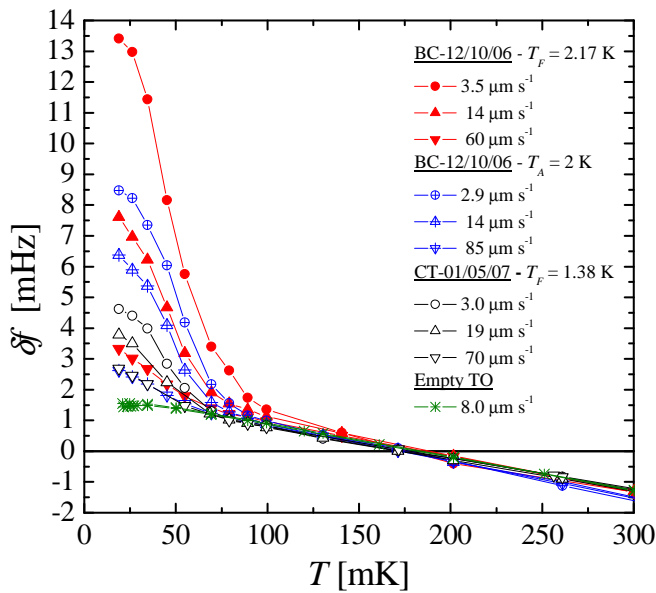}
\caption{\label{fig:one}Temperature dependence of $\delta$\textit{f}
for a BC sample before and after annealing, and one CT sample. For
all samples \textit{x}$_3$ $\approx$ 1 ppb. Data was obtained with
the beryllium-copper TO from Ref.~\onlinecite{tj}. Freezing
(\textit{T$_F$}) and annealing (\textit{T$_A$}) temperatures are
given for each sample. All data was taken during cooling scans in
which the temperature was successively lowered in controlled steps.
For any one sample the measured frequency at \textit{T} $>$
\textit{T$_O$} is identical for all values of \textit{v}$_0$. For
easy comparison we have set $\delta$\textit{f} = 0 at \textit{T} =
175 mK by subtracting 1071.007 Hz, 1071.011 Hz, 1071.035 Hz, and
1071.898 Hz from the data of BC-12/10/06, BC-12/10/06
post-annealing, CT-010507, and the empty TO. Thus, the effective
mass loading for each of these samples is 891 mHz, 887 mHz, and 863
mHz.}
\end{figure}

\begin{figure}[t]
\includegraphics[width=1.0\columnwidth]{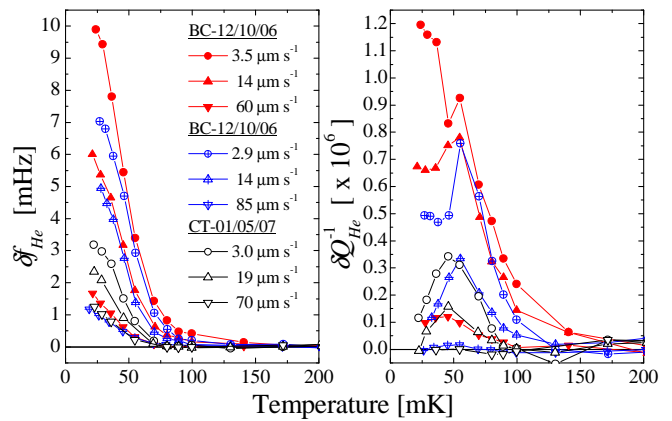}
\caption{\label{fig:two}Temperature dependence of
$\delta$\textit{f$_{He}$} and $\delta$\textit{Q$_{He}$}$^{-1}$ for
the same samples presented in Fig.~\ref{fig:one}, obtained by
subtracting the temperature dependence of the empty cell. The
symbols for BC-12/10/06, pre- and post-annealing, are identical to
those in Fig.~\ref{fig:one}.}
\end{figure}

Sec.~\ref{sec:data} we present our experimental observations in a
general manner by simply describing the dependencies of the observed
frequency shifts and dissipation signals on temperature (see
Figs.~\ref{fig:one} and \ref{fig:two}), oscillation speed
\cite{convenience} (see Fig.~\ref{fig:three}), and growth method.
Two different interpretations of the data are given in
Secs.~\ref{sec:ncri} and \ref{sec:c44}. In the former we attempt to
describe the experiments in the context of a supersolid that is
permeated by vortices. Since vortex lines interact with velocity
fields one of the important parameters in this model is
\textit{v}$_0$. In Sec.~\ref{sec:c44} we discuss the consequences of
attributing the entire frequency shift to an increase in the shear
modulus of solid $^4$He. It is assumed in this picture that the
rigidity of the solid is strongly affected by the motion of
dislocation lines under the applied stress $\sigma$. In anticipation
of these two interpretations, we have simultaneously plotted the
data in Fig.~\ref{fig:three} as a function of \textit{v}$_0$ and
$\sigma$. In addition, in both Figs.~\ref{fig:four} and
\ref{fig:five} we have plotted the frequency shift in terms of NCRIF
and $\delta$\textit{c}$_{44}$/\textit{c}$_{44}$.

\section{\label{sec:data}Experimental results}

A recent study \cite{tj} of samples grown at constant temperature
and pressure (CT/CP) at a single point on the solid-liquid
coexistence curve found the onset of the frequency shift to be sharp
and reproducible, which is consistent with the expected formation of
single crystals. In contrast, wide variations in $\delta$\textit{f}
and \textit{T$_O$} were observed when employing the blocked
capillary (BC) method, which typically yields polycrystalline
samples or highly strained crystals due to the pressure drop during
freezing. In this paper we examine in detail the different thermal
histories of $\delta$\textit{f} in BC and CT/CP samples that follow
from changes in oscillation speed above and below \textit{T$_O$}. We
have measured the temperature dependence of $\delta$\textit{f} in 30
samples, with the bulk of our work being carried out on
``isotopically pure'' (\textit{x}$_3$ $\approx$ 1 ppb of $^3$He)
samples at 1 $\mu$m$\,$s$^{-1}$ $<$ \textit{v}$_0$ $<$ 100
$\mu$m$\,$s$^{-1}$. To overlap with Ref.~\onlinecite{kojima} we also
studied commercially pure (\textit{x}$_3$ $\approx$ 0.3 ppm) samples
up to speeds of 880 $\mu$m$\,$s$^{-1}$.

Figure~\ref{fig:one} shows the temperature dependence of
$\delta$\textit{f} for a BC sample before and after annealing,
compared with that for a sample grown at CT. We find that the
temperature dependence of $\delta$\textit{f} for each sample is
reproducible in warming and cooling scans at low oscillation speeds,
i.e., on the order of one micron per second. Although the
differences between BC and CT/CP samples are apparent at low speeds,
the temperature dependence obtained at \textit{v}$_0$ $>$ 50
$\mu$m$\,$s$^{-1}$ is very similar in all samples studied (including
0.3 ppm samples).

After subtracting the temperature dependence of the empty cell we
obtain the frequency shift and dissipation due to solid $^4$He,
which we respectively denote by $\delta$\textit{f$_{He}$} and
$\delta$\textit{Q$_{He}$}$^{-1}$. These results are plotted in
Fig.~\ref{fig:two}. There is considerable dissipation in rapidly
grown BC samples even at the lowest temperatures. After annealing
sample BC-12/10/06 the high temperature tail of
$\delta$\textit{f$_{He}$} (and thus \textit{T$_O$}), the low
temperature limiting value of $\delta$\textit{f$_{He}$}, and the
magnitude of $\delta$\textit{Q$_{He}$}$^{-1}$ are all reduced. Thus
it appears that annealing affects the temperature dependence of
$\delta$\textit{f$_{He}$} much in the same way as driving the TO at
high speed (see Fig.~\ref{fig:two}). One difference, however, is
that the results from annealing are permanent.

\begin{figure}[t]
\includegraphics[width=1.0\columnwidth]{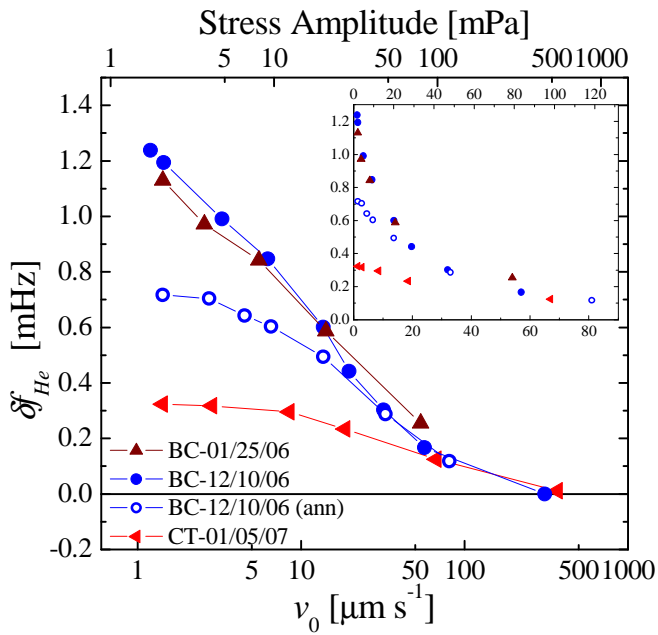}
\caption{\label{fig:three}Dependence of the low temperature value of
$\delta$\textit{f$_{He}$} on oscillation speed and applied stress.
The inertial stress is approximated by $\sigma$ $\approx$
$\pi\rho$\textit{fv}$_0$/2, for a given helium density $\rho$.
Values of \textit{f$_{He}$} were extracted from cooling scans that
began well above \textit{T$_O$}. Samples are labeled according to
growth method and date. The data are plotted versus a linear
abscissa in the inset.}
\end{figure}

In Fig.~\ref{fig:three} we have plotted the low temperature value of
$\delta$\textit{f$_{He}$} measured at different \textit{v}$_0$,
which again demonstrates that the data converge at high speeds. One
interesting difference from previous TO studies
\cite{science,prl,kojima,rr2,kubota} is the lack of saturation in
$\delta$\textit{f$_{He}$} even well below 10 $\mu$m$\,$s$^{-1}$.
This is most evident in two BC samples (01/25/07 and 12/10/06) grown
from the normal fluid phase, where $\delta$\textit{f$_{He}$}
actually doubles when \textit{v}$_0$ is decreased from 10
$\mu$m$\,$s$^{-1}$ to 1 $\mu$m$\,$s$^{-1}$. In samples assumed to be
of higher quality, we find \textit{v$_C$} $\approx$ 3
$\mu$m$\,$s$^{-1}$ (3.5 $\mu$m$\,$s$^{-1}$ corresponds to $\kappa$ =
\textit{h}/\textit{m}).

The curves in Fig.~\ref{fig:three} were compiled from cooling scans
that began well above the onset temperature. Using this procedure we
found that $\delta$\textit{f$_{He}$} is diminished at higher
\textit{v}$_0$ for all \textit{T} $<$ \textit{T$_O$} in a completely
reproducible manner, whereas measurements taken during warming are
history-dependent when obtained at speeds greater than a few microns
per second. As we will show below, varying the oscillation speed
below the onset temperature can result in irreversible changes in
$\delta$\textit{f$_{He}$}. Careful examination of how the
oscillation speed affects $\delta$\textit{f$_{He}$} reveals
metastability at the lowest temperatures, as depicted in
Figs.~\ref{fig:four} and \ref{fig:five} (see Secs.~\ref{sec:ncri}
and \ref{sec:c44}).

Using the following protocol, the thermal history of
$\delta$\textit{f$_{He}$} was investigated in crystals grown under
different conditions. First, the temperature dependence of
$\delta$\textit{f$_{He}$} was measured while cooling at low speed
(\textit{v}$_0$ $<$ 3 $\mu$m$\,$s$^{-1}$). At \textit{T} $\approx$
20 mK \textit{v}$_0$ was slowly increased. Final velocities were
usually $\sim$20 $\mu$m$\,$s$^{-1}$, but in some cases much greater
(e.g., several hundred microns per second). Multiple thermal cycles
were then performed in succession, indicated by the arrows in
Fig.~\ref{fig:four}. Complete equilibration was purposely avoided
while cycling between 30 mK and 60 mK in order to observe multiple
metastable states. The measurement culminated with a high velocity
cooling trace that began well above \textit{T$_O$}.

Figures~\ref{fig:four} and \ref{fig:five} show that for one
oscillation speed a number of different values of
$\delta$\textit{f$_{He}$} ($\propto$ NCRIF and
$\delta$\textit{c}$_{44}$/\textit{c}$_{44}$) can be ``frozen in''
below a characteristic temperature, \textit{T}* $\approx$ 30 mK.
This is true regardless of the sample growth procedure. However,
there are significant differences in the time evolution (following
temperature steps) and the velocity dependence of
$\delta$\textit{f$_{He}$} between samples formed by the BC method
and crystals grown at CT/CP. Similar behavior is seen in 0.3 ppm
samples, but with \textit{T}* $\approx$ 45 mK. For all samples,
there is no observable decay from any of the metastable states of
$\delta$\textit{f$_{He}$} on a timescale of days as long as the
temperature is maintained below \textit{T}*. Further discussion of
the hysteresis is deferred to Secs.~\ref{sec:ncri} and \ref{sec:c44}
below.

\section{\label{sec:ncri}The vortex liquid model}

\begin{figure}[t]
\includegraphics[width=1.0\columnwidth]{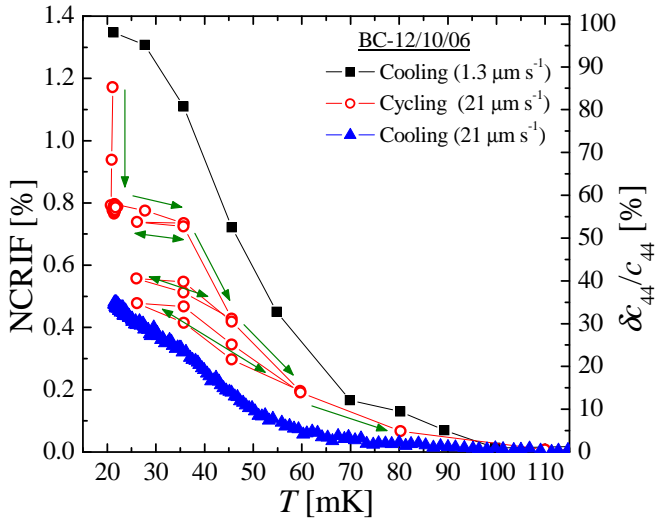}
\caption{\label{fig:four}Metastability of $\delta$\textit{f$_{He}$}
(plotted as NCRIF and $\delta$\textit{c}$_{44}$/\textit{c}$_{44}$)
for a BC sample. The values of
$\delta$\textit{c}$_{44}$/\textit{c}$_{44}$ were calculated based on
the assumption that the solid sample was isotropic, i.e., highly
polycrystalline \cite{vos} due to BC method that was employed. Even
larger values would be obtained if the sample was assumed to be a
single crystal. Decay of the signal occurs for \textit{T} $>$ 30 mK.
Cycling above and below \textit{T}* = 30 mK results in successively
smaller values until equilibrium (i.e., that obtained when cooling
from above \textit{T$_O$}) is reached. The speeds of \textit{v}$_0$
= 1.3 $\mu$m$\,$s$^{-1}$ and 21 $\mu$m$\,$s$^{-1}$ correspond to
$\sigma$ = 2.0 mPa and 32 mPa.}
\end{figure}

\begin{figure}[t]
\includegraphics[width=1.0\columnwidth]{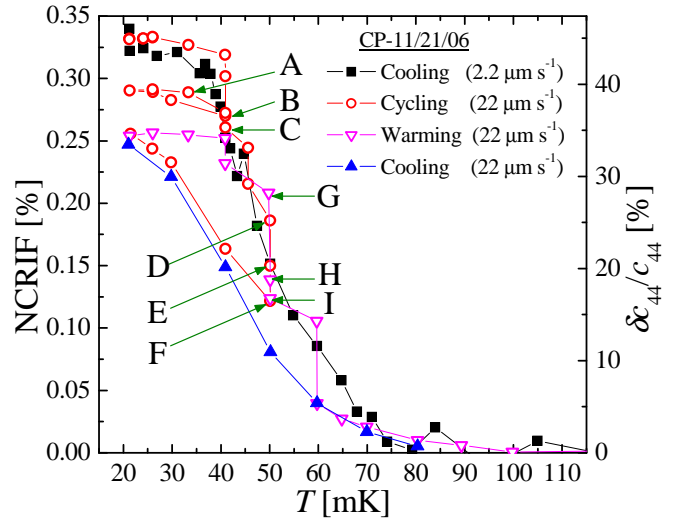}
\caption{\label{fig:five}Metastability of NCRIF and
$\delta$\textit{c}$_{44}$/\textit{c}$_{44}$ for a CP sample. The
sample was assumed to be a single crystal with its c-axis oriented
horizontally. If we instead assume that the c-axis is oriented
vertically (or that the sample is isotropic) the low temperature
limiting value of $\delta$\textit{c}$_{44}$/\textit{c}$_{44}$ would
be 60\% (30\%). Labels A through I denote changes in the decay rate
at several different temperatures. The speeds of \textit{v}$_0$ =
2.2 $\mu$m$\,$s$^{-1}$ and 22 $\mu$m$\,$s$^{-1}$ correspond to
$\sigma$ = 3.4 mPa and 34 mPa.}
\end{figure}

As stated in Sec.~\ref{sec:intro}, the TO experiments to date are
qualitatively consistent with many features of Anderson's vortex
liquid model. The present work, however, demonstrates that below 60
mK the system does not act like a free vortex liquid as originally
proposed. In fact, below 30 mK it exhibits severe pinning. With this
picture in mind we can explain the temperature dependence of NCRIF
($\propto$ $\delta$\textit{f$_{He}$}) in Fig.~\ref{fig:two}, the
velocity dependence in Fig.~\ref{fig:three}, and the history
dependence in Figs.~\ref{fig:four} and \ref{fig:five}.

First we consider the temperature dependence. As the system is
cooled through the transition some number of free vortices are
produced, which is proportional to \textit{v}$_0$. When \textit{T}
$<<$ \textit{T$_O$} the mobility of the existing vortices diminishes
and it is unfavorable to create additional vortex lines. The
inability of vortices to screen supercurrents leads to a sizeable
NCRIF in the low temperature limit. When the system is driven at
high speeds, not only is NCRIF smaller but the high temperature tail
of NCRI is reduced. This suggests that vortex pinning is very weak
at \textit{T} $\ge$ 60 mK. It is also likely that vortices are more
easily excited as \textit{T$_O$} is approached.

The velocity dependence in Fig.~\ref{fig:three} reflects the
distribution in the strength of individual vortex pinning sites at
low temperature. Annealing might effectively remove the weakest
pinning centers so that the low temperature limiting value of NCRIF
is only decreased at small \textit{v}$_0$. At greater speeds these
weak regions are irrelevant since there is a plethora of free
vortices, hence the convergence in Figs.~\ref{fig:two} and
\ref{fig:three} at \textit{v}$_0$ $>$ 50 $\mu$m$\,$s$^{-1}$.

In the context of superconductivity or magnetism (or superfluidity),
the system is prepared differently if (velocity) field-cooling or
zero field-cooling are employed. The hysteresis shown in
Figs.~\ref{fig:four} and \ref{fig:five} is a consequence of what the
authors of Ref.~\onlinecite{kojima} have called vortex glass
behavior. If the state of the system is prepared at low
\textit{v}$_0$ there is only a ``small'' number of vortices frozen
into the solid. When \textit{v}$_0$ is gradually increased while
\textit{T} $<$ \textit{T}*, NCRIF becomes either unstable (see
Fig.~\ref{fig:four}) or metastable (see Fig.~\ref{fig:five}). In the
former case, NCRIF decays until it reaches a metastable state. For
\textit{T} $>$ \textit{T}*, NCRIF is reduced by the enhanced
mobility of vortices.

The low temperature metastability seems to depend on the difference
between the initial NCRIF prepared at low speed and the
``equilibrium'' value obtained upon cooling at the higher speed. For
instance, the difference is NCRIF = 0.9\% for BC-12/10/06 (see
Fig.~\ref{fig:four}) and NCRIF = 0.08\% for CP-11/21/06 (see
Fig.~\ref{fig:five}), resulting in a much more stable value for the
latter. The measured value for BC-12/10/06 immediately drops by
about 40\% of its magnitude following the increase in
\textit{v}$_0$, while there is no change for CP-11/21/06. The
behavior of CP-11/21/06 is similar to that in
Ref.~\onlinecite{kojima}, which was compared with the Meissner
effect (i.e., a robust NCRIF due to the inability of vortices to
enter into the solid). We note that the metastability up to 800
$\mu$m$\,$s$^{-1}$ in Ref.~\onlinecite{kojima} may be related to the
small NCRIF $\approx$ 0.1\% of the sample. In comparison, upon
raising \textit{v}$_0$ to 880 $\mu$m$\,$s$^{-1}$ we observe, for a
0.3 ppm sample with NCRIF $>$ 1\% ($\delta$\textit{f$_{He}$} $>$ 8
mHz), an abrupt drop in the signal that is followed by a gradual
decay. In fact, in this extreme limit of \textit{v}$_0$, the decay
rate is essentially insensitive to temperature steps (warming and
cooling) between 20 mK and 45 mK. Only upon warming above 60 mK does
the decay rate increase further.

\begin{figure}[t]
\includegraphics[width=1.0\columnwidth]{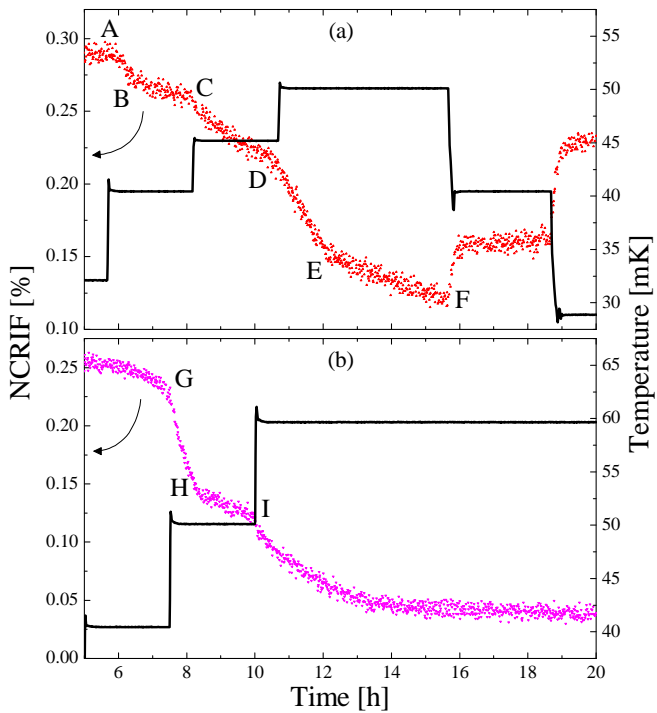}
\caption{\label{fig:six}Segments of raw data used to compile the
``cycling'' and ``warming'' curves in Fig.~\ref{fig:five}. Decay
rates above (D to E and G to H) the low velocity trace are greater
than those below (E to F and H to I). The ordinate axis is given in
terms of NCRIF for easy comparison with Fig. \ref{fig:five}.}
\end{figure}

In 1 ppb samples, the stability is reduced as the temperature is
raised above 30 mK. Fragments of data from scans of CP-11/21/06 are
displayed in Fig.~\ref{fig:six}. The data at any \textit{T} $<$ 60
mK cannot be fit satisfactorily using a simple mathematical formula.
For example, the decay at 40 mK in Fig.~\ref{fig:six}(a) (between
points A and B) begins immediately, but slows down abruptly after
only 1 h. In Fig.~\ref{fig:six}(b), NCRIF barely changes in the
first hour at the same temperature (data prior to point G).
Irregular rates are even more apparent at 50 mK. The decay slows
down dramatically when NCRIF crosses (e.g., points E and H) from the
region above the low velocity cooling trace to the region below it.
This is most obvious in the Fig.~\ref{fig:six}(b), where the sample
was warmed to 50 mK rapidly to result in a larger initial NCRIF. The
very different decay rates in these two regions, and the identical
temperature dependence in warming and cooling scans, both suggest
that there is physical significance to the reproducible data
obtained in the low velocity limit. Similar behavior could not be
observed in BC samples because of higher decay rates, i.e., it was
impossible to exceed the low velocity trace upon warming (see
Fig.~\ref{fig:four}).

Interestingly, the decay in CP-11/21/06 at 60 mK (data following
point I) is different from that at 40 mK and 50 mK in that it can be
well fit to an exponential form, which yields a time constant of
2.25 h. The smooth decay across the low velocity trace may indicate
that the small tail of NCRIF for \textit{T} $\ge$ 60 mK does not
denote the same ``boundary'' as seen at lower temperatures. This
observation is complemented by the enhanced decay rate at 60 mK that
we find in 0.3 ppm samples as well as the \textit{x}$_3$-independent
specific heat peak that has been observed \cite{lin} in the same
vicinity, and suggests that there is some inherent transition
between the ``free'' and ``pinned'' regimes.

\begin{table}[t]
\caption{\label{tab:table1}Approximate stress amplitudes
corresponding to \textit{v$_C$} in various cells. Values are
estimated by ignoring the effects from the top and bottom surfaces
of the cell and equating two different expressions for the torque:
\textit{I}$\alpha$ = \textit{rA}$\sigma$. Here, $\alpha$ is the
angular acceleration and \textit{A} is the area at the rim of the
cell. The stress exerted in an annular geometry differs from that in
a cylindrical one by the approximate factor,
(2\textit{tf})$_{Ann}$/(\textit{rf})$_{Cyl}$.}
\begin{ruledtabular}
\begin{tabular}{cccc}
Cell&\textit{f} [Hz]&\textit{v$_C$} [$\mu$m$\,$s$^{-1}$]&$\sigma$\textit{$_C$} [mPa]\\
\hline
Cylinder\footnotemark[1] & 1072 & $<$ 3.5 & $<$ 5.4\\
Cylinder\footnotemark[2] & 1173 & 15 & 12\\
Cylinder\footnotemark[2] & 496 & 15 & 28\\
Annulus\footnotemark[3] & 912 & 3-40 & 1.6-22\\
Annulus\footnotemark[4] & 912 & 10 & 5.4\\
Annulus\footnotemark[5] & 874 & 40 & 3.3\\
\end{tabular}
\end{ruledtabular}
\footnotetext[1]{Present set of measurements.} \footnotetext[2]{Data
from Ref.~\onlinecite{kojima}.} \footnotetext[3]{Data from
Ref.~\onlinecite{science}.} \footnotetext[4]{Data from
Ref.~\onlinecite{prl}.} \footnotetext[5]{Data from
Ref.~\onlinecite{rr2}.}
\end{table}

Another intriguing result is the nearly equal spacing between each
metastable NCRIF among the four samples that we studied in detail
(e.g., see Figs.~\ref{fig:four} and \ref{fig:five}). It is clear
that the system prefers to possess specific values of NCRIF in the
low temperature limit. For example, counter to the slow decay upon
warming a solid sample, as the temperature is lowered we find NCRIF
to ``jump'' up to a steady value (see the data in Fig.~\ref{fig:six}
following point F). If we naively estimate the frequency shift due
to the presence of a single vortex line using the expression
\cite{hessfetter}, \textit{I$_{Vort}$} =
$\pi\rho$\textit{r}$^4$NCRIF/ln[\textit{r}/\textit{a}$_0$], we
calculate a shift corresponding to 25\% of the observed NCRIF for a
vortex core radius \cite{core} of \textit{a}$_0$ = 0.1 nm. This
value is of the same order of magnitude to what we observe (e.g.,
from Fig.~\ref{fig:five} we get 0.04/0.33 = 11\% for CP-11/21/07).

Although our experimental results can be qualitatively interpreted
within the framework of Anderson's vortex liquid model, very few
quantitative comparisons are possible \cite{penzevnote}. One of the
present experimental results that is difficult to understand in the
vortex picture is depicted in Fig.~\ref{fig:three}. In two samples
the observed frequency shift continues to increase when
\textit{v}$_0$ is reduced to 1 $\mu$m$\,$s$^{-1}$, such that even
the peak to peak amplitude of 2 $\mu$m$\,$s$^{-1}$ is less than that
which corresponds to a single quantum of circulation ($\sim$3.5
$\mu$m$\,$s$^{-1}$).

\section{\label{sec:c44}Anomalous elasticity of solid $^4$He}

Due to the similarities discussed in Sec.~\ref{sec:intro} between
the observed frequency shifts and the results of
Ref.~\onlinecite{day}, we are compelled to interpret the data from
Sec.~\ref{sec:data} in terms of the response of the dislocation
network to oscillating stress fields imposed on the solid.

Here we report our preliminary estimates of changes in the shear
modulus of solid $^4$He. By calculating \cite{FEM} the resonant
frequency of the torsion mode for different \textit{c}$_{44}$ values
we extracted the percentage change in the modulus that corresponds
to a small frequency shift of the TO. The change is given by
$\delta$\textit{c}$_{44}$/\textit{c}$_{44}$ =
(\textit{d}ln\textit{f}/\textit{d}ln\textit{c}$_{44}$)$^{-1}$($\delta$\textit{f$_{He}$}/\textit{f}),
where the derivative term comes from the FEM calculation and
frequency shift comes directly from our TO measurements. The results
of our calculations are shown in Figs.~\ref{fig:four} and
\ref{fig:five}. Since the low temperature change in the resonant
frequency is very small (parts per million), it scales linearly with
the change in the shear modulus. As a result the inferred
temperature dependence of
$\delta$\textit{c}$_{44}$/\textit{c}$_{44}$ is identical to
$\delta$\textit{f$_{He}$}. It is therefore natural that the
qualitative features of the data in Ref.~\onlinecite{day} be very
similar to that from TO studies.

However, the magnitude of the relative change in \textit{c}$_{44}$
appears to be different. In order to account for the observed
$\delta$\textit{f$_{He}$} in TO measurements the required
$\delta$\textit{c}$_{44}$/\textit{c}$_{44}$ values among the samples
studied are between two and five times larger than those seen in
Ref.~\onlinecite{day}. Moreover, the changes are greater than the
theoretical upper limit of 30\% that can result from the pinning of
dislocation lines \cite{soft}. We have carried out similar
calculations for some of the TO's in the literature
\cite{science,prl,kojima,soph} and found that even larger values of
$\delta$\textit{c}$_{44}$/\textit{c}$_{44}$ are necessary to account
for the observed $\delta$\textit{f$_{He}$}. For instance, typical
NCRIF values in Refs.~\onlinecite{science,prl,soph} all translate
into an increase in \textit{c}$_{44}$ by approximately a factor of
10. For the double oscillator from Ref.\onlinecite{kojima} we find
\textit{c}$_{44}$ to increase by factors of 5 and 10 in the
anti-symmetric and symmetric modes, respectively. This implies that
the effect is very dependent on frequency, contrary to the findings
of Day and Beamish. Furthermore, it is difficult to apply the
dislocation pinning picture to the experiments in porous Vycor glass
\cite{nature}, in which the dimensions of helium (i.e., the pore
size) are smaller than typical dislocation loop lengths in bulk
crystals \cite{wanner,iwasa,moredislocations}. This, and the fact
that the crystalline phase \cite{bcc} of the confined $^4$He is
body-centered-cubic rather than hexagonal-close-packed, would lead
one to expect very different behavior in the response to torsional
oscillations, counter to what is observed.

Day and Beamish found the measured shear modulus to be a function of
stress amplitude. Below a critical stress $\sigma$\textit{$_C$} the
magnitude of \textit{c}$_{44}$ saturates at a maximum value, as does
$\delta$\textit{f$_{He}$} in some samples (see
Fig.~\ref{fig:three}). This critical value, $\sigma$\textit{$_C$} =
300 mPa, is smaller than the estimated breakaway stress of 4 Pa for
a $^3$He-$^3$He separation of 5 $\mu$m (a typical loop length
\cite{wanner,iwasa,moredislocations}) along a dislocation, but
closely matches the stress at which $\delta$\textit{f$_{He}$} tends
to zero in Fig.~\ref{fig:three}. Such strong nonlinearity between
the drive and response, also seen in other low frequency acoustic
measurements \cite{yuri}, occurs at ``critical'' speeds that are 100
to 1000 times less (e.g., $\sim$50 nm$\,$s$^{-1}$ in
Ref.~\onlinecite{day}) than in TO studies
\cite{science,prl,rr2,kojima,kubota}. It was concluded in
Ref.~\onlinecite{day} that $\sigma$\textit{$_C$} is a more
fundamental quantity than \textit{v$_C$} due to it being independent
of the measurement frequency. However, this is inconsistent with
Ref.~\onlinecite{kojima} and implies that the two types of
measurements, although clearly related on some level, are distinct.

\section{Conclusions}

We have studied a number of solid $^4$He samples grown within a
torsional oscillator at constant temperature and pressure, as well
as with the blocked capillary method, and found that the resonant
frequency of the system possesses a very strong thermal history. The
magnitude of the apparent NCRIF measured in the low temperature
limit is reproducible when obtained upon cooling the sample from
temperatures well above the transition. Modulation of the
oscillation speed below the onset temperature reveals the existence
of many metastable states.

We find that these properties are qualitatively consistent with two
different interpretations: the response of vortices to velocity
fields and/or the response of dislocation lines within the solid to
oscillating stress fields. The mobility of either of these entities
becomes very limited with decreasing temperature, resulting in
hysteretic behavior below 60 mK. Extreme pinning takes place below
30 mK (1 ppb) or 45 mK (0.3 ppm), such that their motion is
essentially frozen out.

At present there remain discrepancies between the experiments and
either of the two interpretations. In the vortex liquid picture, the
most notable problem is the very low critical velocity found in this
study. In regard to the elastic properties of solid $^4$He, we find
that the majority of the frequency shifts observed are appreciably
larger than upper limit expected from the pinning of dislocations.\\

\begin{acknowledgments}
We thank P. W. Anderson, J. R. Beamish, W. F. Brinkman, D. A. Huse,
J. Jain, H. Kojima, X. Lin, L. Pollet, N. V. Prokof'ev, A. S. C.
Rittner, J. Toner, and J. T. West for their input. We are grateful
to J. A. Lipa for providing us with isotopically pure $^4$He.
Funding was provided by the USA NSF under grants DMR-0207071 and
DMR-0706339.
\end{acknowledgments}

%\bibliography{H2TObib}

\begin{thebibliography}{35}
\bibitem{convenience} We discuss the ``velocity dependence'' in a general sense, in that it
is simply proportional to the drive level input to the system (as is
amplitude, acceleration, stress, etc.).
\bibitem{penzevnote} A very recent attempt at an in depth analysis
has been made in Ref.~\onlinecite{penzev}. If we plot our data in
the same manner as Fig. 3 in Ref.~\onlinecite{penzev}, we observe
very different behavior. At present we do not understand this
discrepancy. However, it is interesting to note that for such plots
we see qualitative differences between CT/CP and BC samples.
\bibitem{soph} Calculations were also done for a TO now being used by Rittner and Reppy. The construction is similar to that used in Ref.~\onlinecite{rr2}, but with a
slightly simpler design that is easier to accurately model.
\end{thebibliography}

\end{document}